\documentclass[12pt,aps,pra,preprint,groupedaddress,superscriptaddress]{revtex4}

\usepackage{graphicx}
\usepackage{amsmath,amssymb,amsfonts}
\usepackage{natbib}

\newcommand{\bd}{\begin{displaymath}}
\newcommand{\ed}{\end{displaymath}}
\newcommand{\be}{\begin{equation}}
\newcommand{\ee}{\end{equation}}
\newcommand{\bs}{\begin{subequations}}
\newcommand{\es}{\end{subequations}}
\newcommand{\ba}{\begin{eqnarray}}
\newcommand{\ea}{\end{eqnarray}}

\begin{document}

\title{Femtosecond response in rare gas matrices doped with NO
impurities: A stochastic approach}

\author{G. Rojas-Lorenzo}
\affiliation{Instituto Superior de Tecnolog\'{\i}as y Ciencias Aplicadas,
Ave.\ Salvador Allende, esq.\ Luaces, 10400 - La Habana, Cuba}

\author{A. S. Sanz}
\affiliation{Instituto de F\'{\i}sica Fundamental,
Consejo Superior de Investigaciones Cient\'{\i}ficas,
Serrano 123, 28006 Madrid, Spain}

\author{J. Rubayo-Soneira}
\affiliation{Instituto Superior de Tecnolog\'{\i}as y Ciencias Aplicadas,
Ave. Salvador Allende, esq.\ Luaces, 10400 - La Habana, Cuba}

\author{S. Miret-Art\'es}
\affiliation{Instituto de F\'{\i}sica Fundamental,
Consejo Superior de Investigaciones Cient\'{\i}ficas,
Serrano 123, 28006 Madrid, Spain}


\begin{abstract}
The femtosecond response of NO-doped rare gas matrices is studied
within a stochastic Langevin theoretical framework.
As is shown, a simple damped harmonic oscillator model can describe
properly the absorption and emission line shapes associated with the
NO ($A^2\Sigma^+ \longleftrightarrow X^2\Pi$) electronic transitions
inside these media as well as the matrix first-solvation shell
response in a process with two timescales, finding a fairly good
agreement with available experimental data.
This approach thus constitutes an alternative and complementary way to
analyze the structural relaxation dynamics of systems in liquids and
solids, leading to a better understanding of the underlying physics.
\end{abstract}




\maketitle


Many molecular systems undergo local structural changes when they are
photoexcited to electronic states with strong electron-phonon
couplings.
These structural changes may lead to an important reorganization of the
surrounding medium \cite{jortner1,jortner2}.
In this regard, due to their simple structural and well-known physical
and thermodynamical properties, pure and doped rare gas matrices
constitute an ideal benchmark to analyze and understand the fundamental
mechanisms of configurational rearrangements in more complex
molecular systems \cite{jortner1,fugol1,fugol2,bondybey}.
Among them, NO-doped rare gas matrices result of particular interest
due to the many conventional (steady-state) \cite{chergui1,chergui1b,%
bohmer,chergui2a,chergui2b,chergui3} and femtosecond pump-probe
\cite{chergui3,jeannin2,jeannin3,jeannin,vigliotti1,vigliotti2}
spectroscopy experimental studies available for these systems, which
reveal the medium dynamical response when the NO molecule is
photoexcited to one of its lowest Rydberg states.

From previous studies in the literature \cite{chergui1,bohmer,%
jeannin2,jeannin}, the dynamics associated with the NO electronic
transition between the ground state ($X$) and one of its lowest Rydberg
excited states (e.g., $A$) can be described as a four-step process
\cite{chergui1,chergui2a,chergui2b} within the {\it configuration
coordinate model in the harmonic approximation} (CCM-HA), which allows
us to relate experimental spectroscopic information to the potential
energy surfaces of the different electronic states of the impurity.
The four-step process is essentially as follows.
In a first step, the absorption of a photon excites the only unpaired
electron of the NO molecule to a Rydberg state.
This provokes an overlapping between the NO Rydberg orbital and the
matrix shells constituted by the neighboring rare gas (Rg) atoms due to
the larger effective size of the new NO excited state with respect to
its ground state.
As a consequence, the atoms of the first solvation shell undergo a
strong repulsion, which displaces them from their equilibrium positions
and, therefore, leads to an increase of the NO-matrix bond lengths.
Because the process is energetically unfavorable, in a second step, the
lattice relaxes around the excited impurity, with the Rg-atoms reaching
a new equilibrium position.
During this process, most of the energy in excess is dissipated through
the matrix, but a small fraction is invested into an expanssion of the
surroundings beyond the first solvation shell.
This latter amount of energy is going to contribute substantially to
the restoring force in the next step.
This basic mechanism consisting of an oscillating radial
expansion/compression of the cage around the NO molecule, from one
equilibrium cage radius to another one, is known as the {\it electronic
bubble formation} \cite{jortner1,jortner2,fugol1,fugol2,chergui1}, and
is also operative in rare gas liquids and clusters \cite{jortner1,%
jortner2,moeller1,moeller2,moeller3,moeller4}.
For instance, molecular dynamics simulations of this bubble formation
have been carried out \cite{jortner3,jortner4,jortner5} to interpret the
absorption and fluorescence spectra of Xe-dopped Ar$_n$ clusters upon
photoexcitation of the Xe dopant to low-$n$ Rydberg states
\cite{moeller1,moeller2,moeller3,moeller4}.
In the case of photoexcited impurities trapped in rare gases and
van der Waals solids, an extensive literature can also be found
\cite{jortner1,fugol1,fugol2,chergui1,chergui1b,goodman1,goodman2,%
jeannin2,jeannin3,bohmer}.
In a third step, Rydberg fluorescence with the emission of a photon
may occur, which (fourth step) brings the system back to its initial
ground state configuration by relaxation, leading to the equilibrium
cage radius associated with the NO ground state.

As infers from the above described absorption/emission process,
the first and third steps are related to energetic aspects of the
transition (steady-state spectroscopy), while the second and fourth
steps are associated with the solvent relaxation dynamics (femtosecond
pump-probe spectroscopy).
In general, all rare gas matrices respond to photoexcitation by
undergoing an ultrafast expansion of the
first solvation shell during the first 100-200~fs (something similar
also happens in the case of the emission process), where almost
all the solute-solvent interaction energy becomes kinetic.
During this expansion, the cage radius reaches a maximum above
$r_{eq}^{(A)}$, the equilibrium cage radius for the NO molecule in its
Rydberg excited state, and then gets back below it (but always above
$r_{eq}^{(X)}$, the equilibrium cage radius for the NO molecule in its
ground state).
The time at which the cage radius reaches this first minimum is the
so-called {\it first recurrence time} $\tau$ \cite{chergui3,jeannin,%
vigliotti1,vigliotti2}.
Due to the energy relaxation process, for any time $t > \tau$, the cage
radius fluctuates with decreasing amplitude around $r_{eq}^{(A)}$,
which is finally reached asymptotically (like a damped motion).

The ultrafast lattice structural dynamics mediated by NO excitation has
been studied in the last years by means of classical molecular dynamics
(CMD) \cite{jimenez1,jimenez2,german1,german2}, obtaining a reasonable
agreement with the experimental data.
Apart from numerical issues (e.g., introduction of periodic boundary
conditions, cutoff radii or smoothing functions to avoid `border
effects'), an important source for deviations between CMD-based
simulations and experiment \cite{chergui2a,chergui2b} is the
unavailability of accurate interatomic potential surfaces describing
the NO($A$)-Rg interaction, where Rg = Ne, Ar, Kr or Xe.
Therefore, the standard procedure to obtain the potential surface is by
considering pairwise potential models and then carrying out simulations
where the parameters defining such models are varied until an optimal
fit to the experimental data is found.
Non-equilibrium and dissipative properties involved in this kind of
processes (e.g., dissipation, relaxation or friction) can also be
studied by means of CMD.
Nevertheless, besides CMD simulations, these properties can also be
extracted alternatively by means of a stochastic approach, which
renders a complementary view of the dynamics underlying the
corresponding experimental findings.

Here, we propose a fully stochastic classical Langevin model (CLM) to
tackle the problem of the analysis and interpretation of the lattice
relaxation dynamics of NO-doped rare gas matrices.
More specifically, making use of the CCM-HA \cite{chergui1,chergui2a,%
chergui2b}, where the pairwise impurity-lattice and lattice-lattice
interactions are substituted by a {\it global} cage harmonic potential,
the structural relaxation of the first solvation shell of the
solid matrix is described in terms of a simple damped harmonic
oscillator (DHO) model, where the energy transfer to/from the
remaining shells of the rare gas matrix (heat bath) is accounted for
by a {\it damping rate} \cite{weiss,breuer} .
In this way, the DHO-CLM used here avoids the problem of considering
explicitly particular many-body potential energy surfaces, this
resulting in a gain of analytical handiness and, therefore, of physical
insight.
Moreover, it also entails a considerable save of computational
time, something very important when one is interested in the global
understanding and fitting procedure of the experimental data rather
than in a detailed dynamics accounting for them.
In this regard, CLM-based approaches have been
successfully used to study surface diffusion \cite{salva,ruth1,%
ruth2,ruth3,ruth4}.
It is worth stressing that, though the Langevin equation employed in
CLMs is seemingly phenomenological, it can be formally derived from
the so-called Caldeira-Leggett Hamiltonian \cite{caldeira1,caldeira2,%
magal,hanggi1,hanggi2,ruth5,ruth6}, very well known in the theory of
open quantum systems \cite{weiss,breuer}.
The Caldeira-Leggett Hamiltonian, introduced in the dynamics of
open classical/quantum systems in 1983, shows that the effects of the
detailed dynamics of a given bath or thermal reservoir (solid, liquid,
etc.) acting on a system can be replaced by an infinite sum of harmonic
oscillators.
The damping acting over the system can be expressed, in general,
as a time-dependent function involving the corresponding harmonic
frequencies and the system-bath coupling constants.
When the damping function becomes Ohmic (i.e., it is described by a
constant spectral function of the frequency), the generalized Langevin
equation (arising from the Caldeira-Leggett Hamiltonian) describing the
system evolution reduces to a standard Langevin equation.

In the problem of NO-doped rare gas matrices, the electron-phonon
coupling during the electron excitation has been estimated from the
experiment \cite{chergui1} and is directly related to the displacements
of the equilibrium configuration coordinate in the ground and excited
states.
On the other hand, the lattice relaxation (first solvation shell)
can be described by a new damping rate which gives us an idea of the
energy loss to the Rg-atom matrix acting as a heat bath via a sort of
{\it effective} phonon-phonon coupling after excitation within the
electronic bubble model.
The dynamics of the cage radius involving the first solvation shell is
coupled to the remaining solvation shells displaying a damped harmonic
oscillation.
As will be shown below, this damping rate introduces a new timescale.

Thus, the NO absorption/emission spectrum is related to the response
of the matrix first solvation shell (i.e., the shell of first nearest
neighbors around the impurity).
In order to describe the dynamical evolution of the first solvation
shell, the CCM-HA is commonly used in the literature \cite{chergui1,%
chergui2a,chergui2b}, with the configuration coordinate being the
{\it matrix cage radius}, $r$.
Within this model, $r$ represents the mean distance between the NO
impurity and each one of the atoms of the first solvation shell, and
an effective harmonic potential (per mass unit),
\be
 U(r) = \frac{1}{2} \ \! \omega^2_0 (r - r_{eq})^2 ,
 \label{eq3b}
\ee
stands for the impurity-matrix interaction.
In (\ref{eq3b}), $\omega_0$ is the effective harmonic frequency
\cite{chergui1,chergui2a,chergui2b} and $r_{eq}$ is the equilibrium
cage radius of the first solvation shell \cite{jimenez2,german1,%
german2}; $r_{eq}^{(A)}$ and $r_{eq}^{(X)}$ will denote the
equilibrium cage radii after absorption and emission, respectively.
Estimates of the harmonic frequency and the cage radius increment,
$\Delta \equiv r_{eq}^{(A)} - r_{eq}^{(X)}$, have been obtained in
the literature for different matrices from a direct analysis of
experimental spectroscopic data by using either the method of
moments \cite{chergui1} or the semiclassical projection method
\cite{chergui2a,chergui2b}.
The so-called Huang-Rhys factors \cite{chergui1} suggest a decreasing
electron-phonon coupling with the rare gas mass forming the solid
matrix.
This can be understood by taking into account the fact that, as the Rg
atom mass increases, the characteristic NO-Rg bond lengths, and also
the Rg-Rg ones, become larger and, therefore, the NO Rydberg orbital
can be better accommodated within the space inside the first solvation
shell.
The values of the effective harmonic frequency in the excited state,
$\omega_0^{(A)}$, as well as the cage radius increment found for NO-Rg
matrices, with Rg = Ne, Ar, Kr and Xe, are given in Table~\ref{tab1}.
As can be noticed, while $\Delta$ decreases monotonically as the NO-Rg
and Rg-Rg bond length increase for $\omega_0$, the variation with the
mass follows the opposite trend.
Notice that an increase of $\omega_0$ with the mass results
counterintuitive, for one would be tempted to think that heavier
particles will oscillate slower than lighter ones.
As will be shown, in our model the $\omega_0$ frequencies, as well
as the corresponding recurrence times, will be modified due to the
presence of the damping rate.

\begin{table}
 \caption{\label{tab1} Effective harmonic frequencies associated
  with the NO in the ground ($\omega_0^{(X)}$) and the Rydberg excited
  ($\omega_0^{(A)}$) states, and increment of the NO-Rg cage radius
  ($\Delta$) for Rg = Ne, Ar, Kr and Xe, within the configuration
  coordinate model in the harmonic approximation.
  For Ne, Ar and Xe, the value of these magnitudes have been taken from
  Ref.~\cite{chergui2b}, while for Kr they are from Ref.~\cite{chergui1}.}
 \begin{ruledtabular}
  \begin{tabular}{c c c c c}
   & Ne & Ar & Kr & Xe \\
   \hline
   $\omega_0^{(X)}$ (ps$^{-1}$) & 1.26 & 1.23 & 1.35 & 1.53 \\
   $\omega_0^{(A)}$ (ps$^{-1}$) & 1.74 & 1.41 & 1.45 & 1.59 \\
   $\Delta$ (\AA) & 0.59 & 0.41 & 0.18 & 0.14
  \end{tabular}
 \end{ruledtabular}
\end{table}

After assuming the CCM-HA, the effects of the energy transfer process
between the first solvation shell  and the remaining shells from the
solid matrix can be described in terms of the
cage radius by means of the Langevin equation associated with
a DHO, as
\be
 \ddot{r}(t) = - \gamma \ \! \dot{r}(t) + F[r(t)] + \xi (t) .
 \label{eq1}
\ee
In this equation, $F = - \nabla U$ is the deterministic force (per mass
unit) acting over the first solvation shell.
The fluctuating force, $\xi (t)$, is described by a Gaussian white
noise, thus satisfying $\langle \xi (t) \rangle = 0$ and
$\langle \xi (t) \xi (t') \rangle = 2\mu\gamma k_B T\delta(t-t')$,
where $\mu$ is the effective mass associated with the first solvation
shell, i.e., its value is 12 times the Rg-atom mass \cite{chergui1,%
chergui2a,chergui2b}, $T$ is the temperature and $k_B$ the Boltzmann
constant.
Furthermore, $\gamma$ is a {\it constant damping rate} related to the
fluctuating force $\xi(t)$ through the {\it fluctuation-dissipation
theorem} \cite{weiss,breuer}.
As can be noticed, Eq.~(\ref{eq1}) is a standard Langevin equation,
which is used instead of its generalized version because the relaxation
time at which the system (the cage radius) approaches equilibrium with
the heat bath is much longer than the correlation time of the bath
fluctuations ({\it Markovian approximation}).
Phenomenological values of $\gamma$ have been used in the literature in
stochastic Langevin theoretical treatments to describe solute-solvent
energy exchange in liquids and solids \cite{adelman1,adelman2,adelman3,%
vigliotti3}, though it is more commonly related to liquids and gases.

In our simulations, 10,000 stochastic trajectories have been run for
each electronic state up to 5~ps, considering time steps of 1~fs.
Initial conditions are assigned like in the corresponding
CMD simulations \cite{jimenez1,jimenez2,german1,german2}.
The initial positions are randomly generated according to a uniform
distribution centered around $r_{eq}^{(X)}$ or $r_{eq}^{(A)}$,
depending on whether emission or absorption are being studied,
respectively.
Regarding the choice of the initial velocities, they follow a standard
Boltzman distribution at a temperature $T$.
The experiments are carried out at $T = 4$~K, where the nuclear
dynamics of the matrix atoms is dominated by the zero-point motion.
Hence, in order to account for this quantum effect properly in CMD
simulations, a higher effective temperature depending on the Debye
frequency of the solid has to be considered \cite{bergsma} (e.g.,
for a solid Ar matrix, the effective temperature corresponding to
$T = 4$~K is about 49~K).
Thus, the initial velocities have to be re-scaled accordingly (i.e.,
taking into account this effective temperature) before starting the
simulation, which we have carried out by using the velocity-Verlet
algorithm \cite{allen} to propagate in time the stochastic
trajectories.
Then, the absorption and emission bands are calculated by constructing
histograms with the energy differences between the initial (ground
state for absorption, excited state for emission) and the final state
(excited state for absorption, ground state for emission).
In particular, we have considered the electronic transition
between the NO ground state and its first lowest Rydberg state,
i.e., $A^2\Sigma^+ \longleftrightarrow X^2\Pi$.
The medium response to the electronic transition is induced by an
instantaneous switching from the ground state potential to the excited
one (making use of the Franck-Condon principle).
Now, in order to reproduce the experimental bandwidths, $\gamma$ is
chosen as the only fitting parameter (remember that in the homologous
CMD simulations, the fitting parameters usually appear in the potential
energy surface models considered), which is varied until the band
widths are fairly reproduced.
The optimal $\gamma$ values obtained from the DHO-CLM are given in
Table~\ref{tab2}.
\begin{table}
 \caption{\label{tab2} Optimal values of $\gamma$ obtained from the
  DHO-CLM.
  This damping rate governs the relaxation dynamics in the ground state
  after emission ($\gamma_X$) or the excited state after absorption
  ($\gamma_A$).}
 \begin{ruledtabular}
  \begin{tabular}{c c c c c}
   & Ne & Ar & Kr & Xe \\
   \hline
   $\gamma_X$ (ps$^{-1}$) & 0.51 & 0.57 & 0.67 & 1.15 \\
   $\gamma_A$ (ps$^{-1}$) & 0.51 & 0.61 & 1.05 & 1.55 \\
  \end{tabular}
 \end{ruledtabular}
\end{table}
These values deviate less than 10\% from the rates obtained through
CMD after computing the approximated time exponential decay of the
total energy autocorrelation function for the impurity,
\be
 \mathcal{C}_E(t) = \langle E(t) E(t - t_0) \rangle
   - \langle E^2(t) \rangle .
\ee
From this expression one can extract the rate of energy transfer
between the first solvation shell and the surrounding bath during the
corresponding relaxation process and, therefore, of the cage radius
damping towards its equilibrium value.
Fortunately, the band widths are not very sensitive to the $\gamma$
rates allowing us a certain degree of flexibility in the fitting
procedure.
As can be seen, $\gamma$ increases from Ne to Xe matrices and is always
larger in the excited than in the ground state.
Physically, this indicates an important energy loss between the first
solvation shell and the medium or heat bath via an {\it effective}
phonon-phonon coupling, which increases with the mass of the species
constituting the solid matrix and, therefore, its rigidity.
Moreover, this energy loss is also larger when the impurity is
photoexcited.
In the experiment and also in CMD simulations, a rapid relaxation of
the photoexcited NO molecules to the medium is observed \cite{jimenez1,%
jimenez2,german1,german2} for Ar, Kr and Xe matrices, while for Ne
matrices the effect is slower and ``collective'' (see below).
Finally, as Eq.~(\ref{eq1}) describes the dynamics of the first
solvation shell, $\gamma^{-1}$ will also provide a timescale for the
solvent relaxation dynamics, which ranges from $\sim$650~fs for NO($A$)
in Xe to $\sim$2,000~fs in Ne.

\begin{figure}
\includegraphics[width=8.5cm]{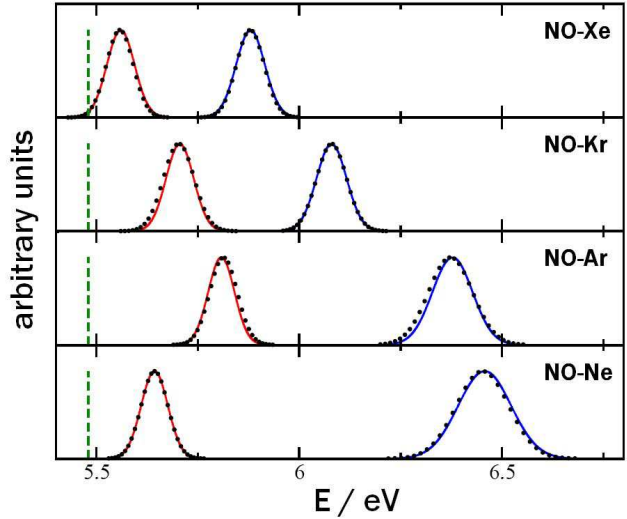}
 \caption{\label{fig1}
  (Color online.)
  Absorption (right/blue) and emission (left/red) line shapes
  associated with the NO ($A^2\Sigma^+ \leftrightarrow X^2\Pi$)
  electronic transition in rare gas solids.
  In all cases, experimental data are represented by dots, while the
  line shapes obtained with the DHO-CLM are denoted by solid lines.
  The green horizontal dashed line around 5.5~eV denotes the position
  of the same absorption/emission process, but for the NO molecule in
  gas phase.}
\end{figure}

In Fig.~\ref{fig1}, absorption (right/blue) and emission (left/red)
bands obtained from the DHO-CLM simulations are plotted for matrices
constituted by different Rg-atoms and compared with the corresponding
experimental data (dots).
These steady-state absorption and emission bands represent the initial
and final stages of the photoinduced structural rearrangement process
around the excited NO molecules.
To compare with the effects produced by the presence of the matrix,
the green horizontal dashed line around 5.5~eV indicates the position
of the NO absorption/emission electronic transition in the gas phase.
As can be noticed, the line shapes of the bands and their Stokes shifts
are in good agreement with the experimental data, the main differences
being lower than 2\% in all cases. Larger numerical discrepancies can
be found, however, in the CMD simulations \cite{jimenez1,jimenez2}.
For example, for Ar matrices, the DHO-CLM reproduces quite accurately
the experimental value of the Stokes shift, 0.570~eV \cite{chergui1},
while the CMD calculations provide a shift of 0.700~eV \cite{jimenez1,%
jimenez2}.
In the case of Ne, the shift obtained from CMD is not so good in
comparison with the experiment and depends strongly on the parameters
of the pairwise potentials considered.
On the other hand, although for Kr and Xe matrices both the DHO-CLM and
CMD \cite{german2} simulations reproduce fairly well the experimental
value of the Stokes shift \cite{chergui1}, the corresponding line
shapes have not been calculated from the latter.

\begin{figure}
\includegraphics[width=8.5cm]{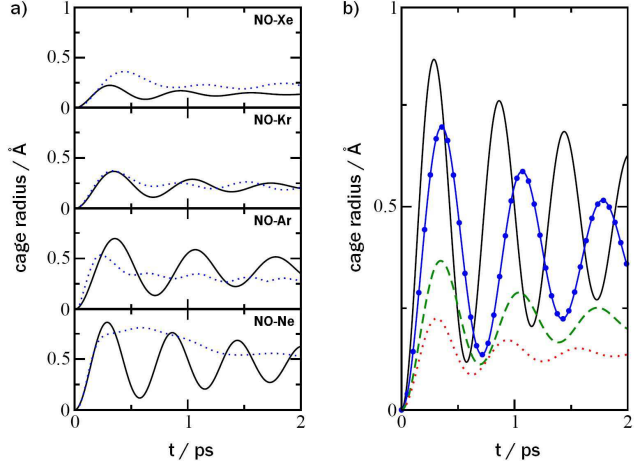}
 \caption{\label{fig2}
  (Color online.)
  Cage radius dynamics.
  a) Blue dotted lines have been obtained from CMD simulations
  \cite{jimenez2,german1,german2}, while black solid lines represent
  the results from the DHO-CLM.
  b) Comparison of the four cases considered in part (a): Ne-matrix
  (black solid line), Ar-matrix (blue solid line with circles),
  Kr-matrix (green dashed line) and Xe-matrix (red dotted line).}
\end{figure}

The time evolution of the cage radius after the absorption is plotted
in Fig.~\ref{fig2}(a), where the cage radius is expressed in terms of
its effective increase, $r' = r - r_{eq}^{(X)}$.
In this way, at $t = 0$, $r' = 0$ means that the cage radius starts at
$r = r_{eq}^{(X)}$; on the contrary, asymptotically, in the limit
$t \to \infty$, $r'$ will approach the cage radius increment $\Delta$,
since the cage radius approaches the equilibrium value, $r_{eq}^{(A)}$.
Physically, this curve describes a process with two different
timescales.
The first part of the process consists of the initial cage radius
expansion (from $t=0$ to the first maximum) and indicates the matrix
response after a photon is absorbed by NO and its unpaired electron is
promoted to the Rydberg state, thus increasing the effective size of
the impurity (and so the size of the bubble).
As seen in Fig.~\ref{fig2}(a), this expansion (solid line) is
relatively fast and mimicking the CMD results (blue dotted line)
for about the first 150~fs in all cases.
In particular, the maximum expansion velocity (slope) reached by the
first solvation shell is 471~m/s (139~fs) for Ne-matrices, 308~m/s
(171~fs) for Ar-matrices, 167~m/s (163~fs) for Kr-matrices, and 114~m/s
(141~fs) for Xe-matrices (between brackets, the time at which this
velocity is reached), which are similar to those obtained from CMD
\cite{jeannin}.
However, because the electron-phonon coupling decreases with
the Rg-atom mass, as derived from the method of moments in
Ref.~\cite{chergui1}, not only a smaller increase of the cage radius
from Ne to Xe matrices is noticeable, but also in the initial slope of
its evolution, as seen in Fig.~\ref{fig2}(b).
The second part of the process describes the relaxation of the cage
radius to its equilibrium value after the initial boosting led by
the (ultrafast) expansion of the NO orbital.
This is the stage that follows the first maximum in the curves
displayed in Fig.~\ref{fig2}(a).
As can be noticed, the damping of the oscillations around the
corresponding equilibrium cage radius increases from Ne to Xe matrices,
this being indicated by an also increasing value of $\gamma$ (which we
have obtained from the fitting to the experimental spectroscopic
measures, as indicated above).
According to the DHO model, the corresponding oscillations displayed in
Fig.~\ref{fig2} and their envelopes follow an expression of
the type $e^{- \gamma t /2} \cos (\bar{\omega}t + \delta)$, where the
renormalized harmonic frequencies are given \cite{weiss} by
\begin{equation}
 \bar{\omega} = \sqrt{\omega_0^2 - \frac{\gamma^2}{4}}
\end{equation}
and the phase shift by $\delta = \tan^{-1} (\gamma/2\bar{\omega})$.
From Ne to Xe matrices, these renormalized frequencies are 1.72, 1.38,
1.35, and 1.39~ps$^{-1}$, respectively. These values follow the same
tendency observed in the experiment \cite{chergui1} by means of an
analysis carried out by the method of moments (see Table~\ref{tab1}).
In principle, a larger damping implies a lower oscillation, which
stresses with the fact that the Rg-atom mass increases.
In other words, intuitively, one should expect that the heavier the
Rg-atom mass, the slower the oscillation and, therefore, the larger
the recurrence times (which are discussed below).
In order to understand this counterintuitive behavior, remember that
bond lengths increase with the Rg-atom mass and, therefore, one could
expect a matrix response more typical of rigid solids for larger Rg
atoms, and more ``floppy'', i.e., involving more neighboring shells,
for smaller Rg atoms.
If so, as happens with sound waves, the transmission of energy through
the material is more effective (faster) for solids than for liquid
which, within the context of our model, means greater energy losses
to the first-neighbors shells as the Rg-atom mass increases.

As shown in Fig.~\ref{fig2}(a), for Ne matrices the DHO-CLM renders
a value for the first recurrence time of $\tau_{\rm DHO-CLM} \approx
564$~fs, which does not agree with the experimental value
$\tau_{\rm exp} \approx 1.1$-1.4~ps \cite{vigliotti1,vigliotti2}.
In the case of CMD, in order to obtain a good estimate of this time,
an artificially shortened Born-Mayer potential model was considered,
which provides $\tau_{\rm CMD} \approx 1.3$~ps \cite{vigliotti1,%
german1}.
To understand this discrepancy, as mentioned above, note that the
NO($X$)-Ne and Ne-Ne bond lengths are smaller than the NO($A$)-Ne one,
which covers about four solvation shells \cite{vigliotti1,german1}.
Thus, although the initial expansion of the cage radius is well
described by the DHO-CLM because it is only affected by the first
solvation shell, immediately after such an expansion the size of the
excited NO wave function overlaps the matrix atoms of upper shells,
which are not considered in the DHO-CLM.
The contribution of these upper shells would lead to observe the
first recurrence at longer times due to a slower relaxation (i.e., a
``floppy''-like matrix behavior under a local structural perturbation).
The recurrence time, therefore, would not be ruled by one shell but by
four shells at a time, as shown through CMD \cite{vigliotti1,german1}.
A simple way to deal with this effect consists of including an
anharmonic contribution in the effective potential acting over
the solvation shell in the involved excited state.
This procedure improves the value of the recurrence time to be
$\tau_{\rm DHO-CLM} \approx 1.09$~ps and leaves unchanged the
corresponding line shapes.
Obviously, this is the simplest and fastest way of considering the
inclusion of more external shells.
Other more elaborated schemes can be envisaged such as, for example,
using a four-variable model with four coupled Langevin equations, each
one describing the evolution of the radius associated with each shell.
In the case of Ar matrices, we find a recurrence time greater than in the corresponding CMD calculation [see Fig.~\ref{fig2}(a)],
 $\tau_{\rm CMD}
\approx 500$~fs \cite{jeannin} versus $\tau_{\rm DHO-CLM} \approx
711$~fs, in better agreement with the experimental value $\tau_{\rm exp}
\approx 600$-800~fs \cite{jeannin,vigliotti2}.
Contrary to the case of Ne, here the NO($A$)-Ar bond length is such
that the expansion of the NO wave function up to $t \sim \tau$ is going
to affect essentially the first solvation shell and, therefore, the
dynamics will be well reproduced.
Note that, as $\mu$ increases (from Ne to Xe), the amplitude of the
expansion curve of the cage radius decreases as well as
the expansion velocity of the solvent shell.
However, the first recurrence time is not an increasing monotomic
function of the mass, as observes in Fig.~\ref{fig2}(b).
With respect to Kr matrices, we have found an acceptable similarity in
the response between DHO-CLM and CMD [note in Fig.~\ref{fig2}(a) that
$\tau_{\rm DHO-CLM} \approx \tau_{\rm CMD}$].
For Xe matrices, both simulations provide a similar behavior pattern
qualitatively (regarding the amplitude of the oscillations) though some
quantitative discrepancies can be found.
To our knowledge, there are no reported femtosecond experiments for the
bubble dynamics in Kr and Xe matrices and, therefore, the corresponding
curves (as also happens with those obtained from CMD) have to be
considered as predictive.

Summarizing, we have shown that the experimental absorption/emission
bands associated with the $A^2\Sigma^+ \leftrightarrow X^2\Pi$
electronic transition of NO in rare gas matrices can be fairly well
reproduced by means of a simple damped harmonic oscillator model, which
describes the dynamics of the first solvation shell (i.e., the response
of the matrix to the NO excitation) in a process with two different
timescales.
The first part of the process (fast expansion of the first solvation
shell) is directly related to the electron-phonon coupling which has
measured in the experiment.
In the second part of the process, the introduction of a new damping
rate via an {\it effective} phonon-phonon coupling after excitation
in order to describe the lattice relaxation allows us to obtain very
reliable band structures by only using data from spectroscopic
measurements and a single fitting parameter ($\gamma$).
As shown, the expansion dynamics of the cage radius is well described
for about the first 200~fs as well as the first recurrence time.
Our DHO-CLM should improve with the mass of the matrix atoms.
Note that the calculation of $\tau$ from the model presented here
marks the time at which a {\it well-defined} first recurrence takes
place.
Nonetheless, the fact that only the first solvation shell is involved
in the dynamics described restricts the applicability of the model
when `collective' shell behaviors are present, as in the case of Ne
matrices.
In such cases, it has been shown that the addition of anharmonic
contributions to the effective harmonic potential used, which would
account for the eventual distortions undergone by the first solvation
shell, render better recurrence times in comparison with the
experimental ones.
For Ar matrices, $\tau_{\rm DHO-CLM}$ agrees fairly well with the
experimental value without any need to consider anharmonicities.
The oscillations of the cage radius as well as the recurrence times
are easily explained within the DHO model through a renormalization
of the harmonic frequencies due to the presence of the damping rate.
This information could be therefore used to design new femtosecond
pump-probe experiments.
The fact that all of these features are reasonably well reproduced by
simple, stochastic Langevin simulations  opens up a new theoretical
way to explore, describe and understand the physics underlying
rearrangement processes in solids, stressing the insightful value
of the model.


We would like to acknowledge fruitful discussions with Prof.\ Majed
Chergui.
We also highly appreciate the very constructive and positive comments
made by one of the referees.
This work has been supported by the Ministerio de Ciencia e
Innovaci\'on (Spain) under Projects FIS2007-62006 and SB2006-0011.
A.S.\ Sanz acknowledges the Consejo Superior de Investigaciones
Cient\'{\i}ficas for a JAE-Doc Contract.


\end{document}